\documentclass{epl}
\title{Reply to ``Comment on Direct equivalence between quantum phase
transition phenomena in radiation-matter and magnetic systems:
scaling of entanglement"}
\shorttitle{Direct equivalence between quantum phase
transition etc.}
\author{J. Reslen\inst{1}\thanks{E-mail:\email{j-reslen@uniandes.edu.co}} \and L. Quiroga\thanks{E-mail:\email{lquiroga@uniandes.edu.co}} \inst{1} \and N. F. Johnson\inst{2}\thanks{E-mail:\email{n.johnson@physics.ox.ac.uk}}}
\institute{
  \inst{1} Departamento de F\'{\i}sica, Universidad de Los Andes, A.A.4976,
Bogot\'a D.C., Colombia \\
  \inst{2} Centre for Quantum Computation and Department of Physics,
University of Oxford, Clarendon Laboratory, Parks Road, OX1 3PU, U.K
}
\shortauthor{J. Reslen \et al.}

\begin{document}

\maketitle

Our Letter \cite{reslen} shows that there is an {\em equivalence} between the Dicke model and the
infinitely range XX (Lipkin) model, at the level of the effective Hamiltonian describing the qubit
reduced subsystem. The Comment by Brankov et al. asserts that our use of a reduced Hamiltonian in
place of the original Dicke Hamiltonian, is ``wrong''  as compared to the results which they previously reported.

Brankov et al.'s criticism is misguided -- in short, they are wrong to use the word ``wrong" since
we make no statement, anywhere in our Letter, to claim that our effective (i.e. reduced) qubit
Hamiltonian is exact at finite temperatures. Yet it is on this basis, and this basis alone,
that Brankov et al. use the word ``wrong". There is nothing wrong with our Letter. As a side note,
we emphasize that their own earlier work is effectively irrelevant to the core results of our paper.
In addition, there are several other confused statements in the Comment by Brankov et al. which,
for the sake of completeness, we point out below.

First, the goal of our Letter was to study the entanglement features of the qubit subsystem, {\em not}
the entire qubit-radiation system. We focused on the quantum phase transition which, by definition,
occurs at {\em zero temperature}. In particular, we looked at the finite-size scaling of two-qubit
entanglement and contrasted it with previously published numerical results for the full Dicke
Hamiltonian \cite{brandes1}. Our main finding was that even at the level of the lowest-order
approximation (i.e. summing special parts of {\it all cumulants}) our reduced effective Hamiltonian
at zero temperature yields a quantum phase transition at the {\em same} critical coupling constant
as the full Dicke Hamiltonian. Even more importantly, our reduced effective Hamiltonian shows
the {\em same} critical finite-size scaling exponents for two-qubit entanglement (i.e. concurrence).
In addition, we found a universal scaling function for the behavior of the concurrence.

Second, at no point in our Letter do we dispute Brankov et al.'s
claim that our {\em temperature-dependent} Hamiltonian (Eq.(8) in
\cite{reslen}) is not exact. Obviously it is not. Instead, and as
stated, it is the result of a lowest-order approximation. In
short, we never presented it as an exact result for non-zero
temperatures. However, what is remarkable about our result is that
at zero temperature it does coincide {\em perfectly} with the
exact one, and also with the finite-size quantities which can be
obtained using it. This is acknowledged by Brankov et al. in their
Comment, hence they do agree with our results. We emphasize that
although the cumulant approach which we used \cite{becker} can be
worked out to any order, the expressions of higher order terms
turn out to be analytically quite complex -- hence our finding
that lowest-orders can be quantitatively accurate at zero
temperature, is very important.

Finally, our work is the first to have shown explicitly that the critical scaling exponents of
two-qubit entanglement (i.e. concurrence) are indeed identical for two different physical models
(i.e. Dicke and Lipkin models). In addition, we showed that the quantum phase transitions of both
models belong to the same universality class. Brankov et al. do not dispute this finding, nor there
is any proof in their Comment to invalidate our basic result. The question raised concerning the
critical behaviour of finite-size systems remains an open problem of course -- but this was beyond
the scope of our Letter.

This work was supported by
the Faculty of Sciences (U. de los Andes), COLCIENCIAS (Colombia) project
1204-05-13614 and DTI-LINK (UK).

\end{document}